\documentclass[reprint,sort&compress,aps,twocolumn,prb,superscriptaddress]{revtex4-2}
\usepackage{csquotes}
\usepackage{graphicx}
\usepackage{float}
\usepackage{dcolumn}
\usepackage{color}
\usepackage{physics}
\usepackage{bm}% bold math
\usepackage[T1]{fontenc}
\usepackage{hyperref}
\usepackage[normalem]{ulem}
\usepackage{color}
\usepackage{hyperref}
\usepackage{dsfont}
\hypersetup{
     colorlinks   = true,
     citecolor    = blue
}

\begin{document}
\newcommand{\mos}{MoS$_2$}
\preprint{APS/123-QED}

%\title{Electronic properties of twisted bilayer graphene - hexagonal Boron Nitride van der Waals heterostructures}
%\title{Modeling the electronic properties of twisted bilayer graphene}
%\title{Atomistic calculations of twisted bilayer graphene and hexagonal Boron Nitride heterostructures}

\title{An accurate description of the structural and electronic properties of twisted bilayer graphene-boron nitride heterostructures}

%\title{\ttfamily HARRY POTTER \\ and the \\ Twisted bilayer graphene/boron nitride heterostructures: An accurate description of their structural and electronic properties}

%Competing interlayer interactions and substrate effects in twisted bilayer graphene on hexagonal Boron Nitride

\author{Min Long}
\affiliation{Key Laboratory of Artificial Micro- and Nano-structures of Ministry of Education and School of Physics and Technology, Wuhan University, Wuhan 430072, China}
\author{Pierre A. Pantale\'on}
\affiliation{Imdea Nanoscience, C/ Faraday 9, 28015 Madrid, Spain}
\author{Zhen Zhan}
\email{zhen.zhan@whu.edu.cn}
\affiliation{Key Laboratory of Artificial Micro- and Nano-structures of Ministry of Education and School 
of Physics and Technology, Wuhan University, Wuhan 430072, China}
%\affiliation{Donostia International Physics Center (DIPC)–UPV/EHU, E-20018 San Sebastián, Spain}
\author{Francisco Guinea}
\affiliation{Imdea Nanoscience, C/ Faraday 9, 28015 Madrid, Spain}
\affiliation{Donostia International Physics Center, Paseo Manuel de Lardiz\'{a}bal 4, 20018 San Sebastián, Spain}
\affiliation{Ikerbasque. Basque Foundation for Science. 48009 Bilbao. Spain.}
\author{Jose \'{A}ngel Silva-Guill\'{e}n}
\affiliation{Imdea Nanoscience, C/ Faraday 9, 28015 Madrid, Spain}
\author{Shengjun Yuan}
\email{s.yuan@whu.edu.cn}
\affiliation{Key Laboratory of Artificial Micro- and Nano-structures of Ministry of Education 
and School of Physics and Technology, Wuhan University, Wuhan 430072, China}

%\date{\today}
%%%%%%%%%%%%%%%%%%%%%%%%%%%%%%%%%%%%%%%%%%%%%%%%%%%%%%%%%%%%%%%%%%%%%%%%%%%%%%%%
\begin{abstract}
Twisted bilayer graphene (TBG) has taken the spotlight in the condensed matter community since the discovery of correlated phases at the so-called magic angle. 
Interestingly, the role of a substrate on the electronic properties of TBG has not been completely elucidated. 
Up to now, most of the theoretical works carried out in order to understand this effect have been done using continuum models. 
In this work, we have gone one step ahead and have studied heterostructures of TBG and hBN using an atomistic tight-binding model together with semi-classical molecular dynamics to take into account relaxation effects. 
We found that the presence of the hBN substrate has significant effects to the band structure of TBG even in the case where TBG and hBN are not aligned.  Specifically, the substrate induces a large mass gap and strong pseudomagnetic fields which break the layer degeneracy.  Interestingly, such degeneracy can be recovered with a second hBN layer. Finally, we have also developed a continuum model that describes the tight-binding band structure. Our results show that a real-space tight-binding model in combination with semi-classical molecular dynamics are a powerful tool to study the electronic properties of supermoir\'e systems and that using this real-space methodology could be key in order to explain certain experimental results in which the effect of the substrate plays an important role. 
\end{abstract}

\maketitle

%%%%%%%%%%%%%%%%%%%%%%%%%%%%%%%%%%%%%%%%%%%%%%%%
\section{Introduction}

Graphene has brought a lot of excitement to the scientific community since its isolation in 2004~\cite{Novoselov2004,CastroNeto2009,Roldan2017}.
Although the properties of twisted bilayer graphene (TBG) have been studied in the past decades~\cite{SuarezMorell2010,Bistritzer2011}, recently it has become of great interest due to the finding of highly correlated phases such as superconductivity and Mott insulating phases when the twist angle is in the so-called magic-angle regime~\cite{cao2018unconventional,cao2018correlated}.
%Furthermore, later works have shown the appearance of insulating and ferromagnetic phases which have made TBG a more appealing material to study.
%In the latter state, an anomalous quantum hall effect (AQHE) has been measured and the material has been thought to be a Chern insulator\cite{Setal20,Sharpe2021orbital}.
%Talk something about the correlated phases etc??

TBG is usually supported on top of a variety of substrates in experiment.  
Interestingly, depending on the experimental setup, it can also be clamped between two layers of a material. 
Among all of these, due to its atomically smooth surface that is relatively free of dangling bonds and charge traps \cite{dean2010boron}, hexagonal boron nitride (hBN) stands out due to the cleaner structures that can be fabricated.
%is widely in order to have
In fact, it has been shown that graphene mobility, when deposited on hBN~\cite{dean2010boron,gannett2011boron}, is dramatically improved as compared to those on SiO$_2$ substrates~\cite{kretinin2014electronic,tan2014electronic}. Furthermore, the large band gap of hBN has the advantage of having little interaction between the states of hBN and graphene which is opposite to the case with a metallic substrate where a large hybridization between the graphene and metal states can occur~\cite{2008giovannetti,2009khomyakov}.

Nevertheless, although graphene supported on hBN has a small interaction due to the large band gap of hBN,~\cite{cao2018correlated} in the case of TBG on top of hBN it has been shown that this small interaction can affect the electronic properties of TBG. 
Interestingly, the appearance of superconducting, insulating and ferromagnetic phases in TBG supported on a nearly aligned hBN substrate, have made this system even more appealing to study~\cite{Moriyama2019,Serlin2019,Setal20,Sharpe2021orbital}.
In the latter state, an anomalous quantum hall effect (AQHE) has been measured and the material has been thought to be a Chern insulator~\cite{Setal20,Sharpe2021orbital}.
Moreover, a ferroelectric phase has been found in Bernal-stacked bilayer graphene sandwiched between two hBN layers~\cite{zheng2020unconventional}.
Therefore, the effect of the substrate supporting or embedding TBG could play a key role on the electronic properties of the system.
%a not completely aligned
%Talk something about the correlated phases etc??

From a theoretical point of view, TBG has been studied using full or effective tight-binding (TB) models as well as continuum models due to the large number of atoms found in the moir\'e supercell~\cite{Koshino2018a,Guinea2019,Guinea2019CModels}.
On the other hand, due to the extra layer added when investigating the properties of TBG supported on hBN, the TB approach is hindered by the large number of atoms and most studies use continuum models.
% continuum models are the more suitable approach
Regarding the effect of the structural relaxation on the electronic properties of the system, it is treated in a perturbative way, as well as is the case of the interaction of TBG with the substrate.
Consequently, although in some works the relaxation of the lattices and the effect of the substrate are taken into account, the resulting model could be in some cases oversimplified and a more detailed calculation that takes into account a realistic relaxation and electronic structure is fundamental in order to explain certain experimental results, as well as to have a good starting point to obtain a good set of parameters for the continuum model~\cite{Shi2021Conm,Wolf2018}.

Therefore, to achieve this realistic model we have taken another approach to the problem. 
We have developed a full tight-binding model for TBG where we also include the effect of hBN in an atomistic level. 
Moreover, to include the effect of the relaxation we have performed semi-classical molecular dynamics in the systems that we study. Finally, we have developed a continuum model that correctly describes the TB bands and that takes into account the relaxation effects as well as the effect of hBN.

This paper is organized in the following way: 
In section \ref{sec:methods} we explain in detail the methodology behind our calculations. In section \ref{sec:subs-ind-pot} we study the potentials that are induced in TBG due to the incorporation of a substrate. Next, we show the main results for the electronic properties of the different structures studied, that is TBG on a hBN substrate and TBG embedded between two layers of hBN (section \ref{sec:results}). We then move to the description of the continuum model in section \ref{sec:cont-mod}. Finally, we give some general conclusions of this work.

\section{Model and Methods}\label{sec:methods}
\subsection{Geometry description}
\begin{figure}
\centering
\includegraphics[width=0.43\textwidth]{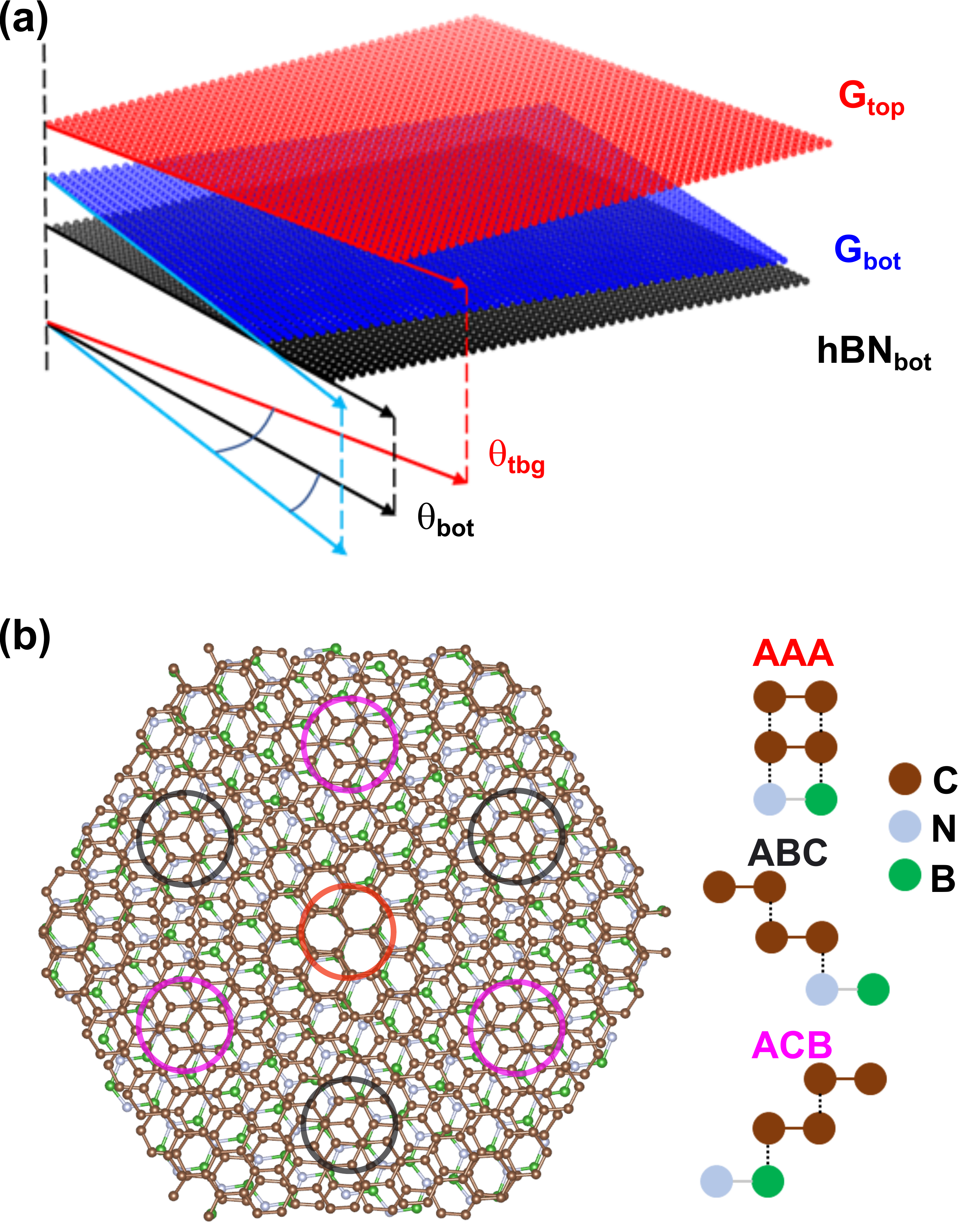}
\caption{(a) Schematics of the atomic configuration of TBG/hBN. (b) Top view of the atomic configuration of TBG/hBN. High-symmetry stacking regions of AAA, ABC and ACB are marked by red, black and purple circles.
Carbon, boron and nitrogen atoms are depicted by brown, green and gray, respectively.
%We select the bottom graphene layer as a reference layer.
}\label{fig:atomic}
\end{figure}

In this work we study two van der Waals heterostructures: a trilayer system composed of TBG lying on top of a hBN layer (TBG/hBN) and a tetralayer system, where TBG is encapsulated by two layers of hBN (hBN/TBG/hBN).  
In the TBG/hBN case, as shown in Fig.~\ref{fig:atomic}, we define the stacking geometry by starting from a non-rotated AAA configuration, where the AA sites of TBG share the same in-plane position $(x,y)=0$ with the nitrogen atoms of hBN  and, moreover, the graphene and hBN bonds are parallel. 
We then rotate the hBN layer, $(hBN_{bot})$, and the top graphene layer, $(G_{top})$, around the origin while keeping fixed the middle graphene layer, $(G_{bot})$.  As schematically shown in Fig.~\ref{fig:atomic}(a), the twist angles are given by $\theta_{tbg}$ for the angle between graphene layers and $\theta_{bot}$ for the angle between hBN and the lower graphene layer.  
In the tetralayer structure, Fig.~\ref{fig:sandwich}, the top hBN layer is rotated by an angle $\theta_{top}$ with respect to $G_{bot}$. We clarify that all twist angles (unless specified) are measured with respect to $G_{bot}$.

We define the lattice vectors of the hexagonal lattice as $a_1=a (\sqrt{3}/2,1/2)$ and $a_2=a (\sqrt{3}/2,-1/2)$ with $a$ the lattice constant. For graphene $a_g=0.246$ nm, and for hBN $a_{hBN}=0.2503$ nm.  As detailed in section A of the supplentary material (SM), the twist angle of TBG can be solely determined by a coprime integer pair $(n,m)$ given by
\begin{equation}
    \theta_{tbg}=2\mathrm{arcsin} \left(\frac{1}{2\sqrt{m^2+mn+n^2}}\right), \label{equa: angletbg}
\end{equation}
with the moir\'e length given by
\begin{equation}
    L_{TBG}=a_g\sqrt{m^2+mn+n^2}=\frac{a_g}{2|\sin{\theta_{tbg}/2}|}.
    \label{eq: Ltbg}
    \end{equation}
Similarly, for the supercell formed by the bottom graphene layer and hBN we have,
\begin{equation}
    L_{hBN}=\frac{(1+\delta)a_g}{\sqrt{\delta^2+2(1+\delta)(1-\cos{\theta_{bot}})}},
    \label{eq: LhBN}
\end{equation}
where $\theta_{bot}$ is the twist angle between the bottom layer graphene and bottom hBN, and the lattice mismatch is $\delta=(a_{hBN-a_g})/a_g$.
A commensurate structure of TBG/hBN structure is achieved when $L_{TBG}=q L_{hBN}$, with $q$ an integer. For example, the structure with $\theta_{bot}=0.53^\circ$, $1.59^\circ$, $2.65^\circ$ in Fig. \ref{fig:potential_band} have $q=1$, 3, 5, respectively. In some cases, even after choosing an appropriate value of $q$, we needed to add a small value of strain to the hBN layer to make the structure commensurate. Here, it is important to note that by properly choosing the values of $\theta_{top}$, $\theta_{bot}$ and $\theta_{tbg}$, the strain can be smaller than $0.1\%$.  Such a small value of strain will not affect the structural or electronic properties of hBN, which, therefore, will not change the properties of the TB/hBN systems that we study in this work. For the interlayer distances, we choose as a starting point the same value for the separation  between the graphene and hBN and the two graphene layers, that is, $d_0=0.335$ nm~\cite{slotman2015effect}.

%From Eqs.~\ref{eq: LhBN} and ~\ref{eq: Ltbg} the commensurability condition is fulfilled when the supercell is the same for both TBG and hBN. 
%Notice that this can be achieved if $L_{hBN}=q L_{tBG}$ with $q$ an integer. To fix an integer $q$ in order to reach a commensurate condition, we added an small strain on the hBN layer. We have found that by properly choosing $\theta_{top}$, $\theta_{bot}$ and $\theta_{tbg}$ the strain value can be smaller than $0.1\%$. In the following, we assume that the interlayer distance between graphene and the nearest hBN layer is $d_{G/hBN}=0.335$ nm, we use the same value for the distance between adjacent graphene layers~\cite{slotman2015effect}.

\subsection{Atomic relaxation}

After constructing a commensurate supercell, to investigate the effect of the hBN substrate on the electronic properties of the system, we fully (both in-plane and out-of-plane) relax the twisted bilayer graphene while keeping the hBN layer fixed in a flat configuration to mimick a bulk or a few layers substrate. 
To do so, we use semi-classical molecular dynamics as implemented in LAMMPS~\cite{plimpton1995fast}. 
For intralayer interactions in the graphene layers, we use the reactive empirical bond order (REBO) potential ~\cite{brenner2002second}. For interlayer interactions between graphene layers, we use the registry-dependent Kolmogorov-Crespi (RDKC) potential developed for graphitic systems~\cite{kolmogorov2005registry}. 
We use the same RDKC potential for the interlayer interaction between graphene and hBN layers. The interactions strength of $C-B$ and $C-N$ are $60\%$ and $200\%$ with respect to the original $C-C$ interaction, respectively~\cite{slotman2015effect}. 
%Both top and bottom hBN layers are kept rigid, mimicking a bulk or a few layers substrate. 
We assume that the relaxed structures keep the periodicity of the rigid cases. 

\subsection{Tight-binding model}

To calculate the electronic properties we use a full tight-binding model given by~\cite{Moon2014} 
\begin{align}
\label{eq: TBHamil}
    \mathbf H&=-\sum_{i,j}t(\mathbf R_i- \mathbf R_j)\ket{\mathbf R_i}\bra{\mathbf R_j}+\sum_i \varepsilon(\mathbf R_i)\ket{\mathbf R_i}\bra{\mathbf R_i}\\\nonumber
    &\qquad+\sum_i V_{D,i}(\mathbf R_i)\ket{\mathbf R_i}\bra{\mathbf R_i}, 
\end{align}
where $\mathbf R_i$ and $\ket{\mathbf R_i}$ represent the lattice position and atomic state at site $i$, respectively, $t(\mathbf R_i-\mathbf R_j)$ is the transfer integral between sites $i$ and $j$. 
The last two terms in the above equation take into account the on-site contributions, where $\varepsilon(\mathbf R_i)$ encodes the carbon, boron and nitrogen on-site energies and $V_{D}(\mathbf R_i)$ the deformation potential resulting from the structural relaxation~\cite{slotman2015effect}.
We assume that $\varepsilon_B=3.34$ eV and $\varepsilon_N=-1.40$ eV, for boron and nitrogen atoms, respectively, and $\varepsilon_C=0$ for carbon atoms~\cite{Moon2014}. The deformation potential will be explicitly described in Sec. \ref{sec:subs-ind-pot}.
For the transfer integral, we simply adopt the common Slater-Koster-type function for any combination of atomic species~\cite{Moon2014}:
\begin{equation}
    -t(\mathbf R)=V_{pp\pi} \left[ 1-\left(\frac{\mathbf R\cdot \mathbf e_z}{R}\right)^2  \right]+V_{pp\sigma}\left(\frac{\mathbf R\cdot \mathbf e_z}{R}\right)^2,
    \label{SK_para}
\end{equation}
with:
\begin{align}
    V_{pp\pi}&=V_{pp\pi}^0 exp\left(-\frac{R-a_0}{r_0}\right), \\   
    V_{pp\sigma}&=V_{pp\sigma}^0 exp\left(-\frac{R-d_0}{r_0}\right),
\end{align}
where $\mathbf e_z$ is the unit vector perpendicular to the graphene plane, $R=|\mathbf R|$, $a_0=a_g/\sqrt{3} \approx 0.142$ nm is the nearest neighbor distance of graphene. 
$V_{pp\pi}^0$ is the transfer
integral between nearest neighbor atoms of graphene and $V_{pp\sigma}^0$ is that of vertically located atoms on the neighboring layers. We take $V_{pp\pi}^0 \approx -2.7$ eV, $V_{pp\sigma}^0 \approx 0.48$ eV. 
We adopt the same Slater-Koster parameters for the interactions between graphene and hBN, and intralayer interactions of hBN \cite{Moon2014}. 
$r_0$ is the decay length of the transfer integral, and is chosen as $r_0 =0.184a$ so that the next nearest intralayer coupling becomes $0.1V_{pp\pi}^0$. We set the cutoff distance of this hopping function to $0.6$ nm since for larger distances the value of hopping energy is small enough that it can be safely neglected. 
We directly diagonalize the Hamiltonian to get the band structure, and compute the density of states of this system using a tight-binding propagation method (TBPM), as described in Ref.~\cite{yuan2010modeling}.

\begin{figure*}[!ht]
    \centering
    \includegraphics[width=\textwidth]{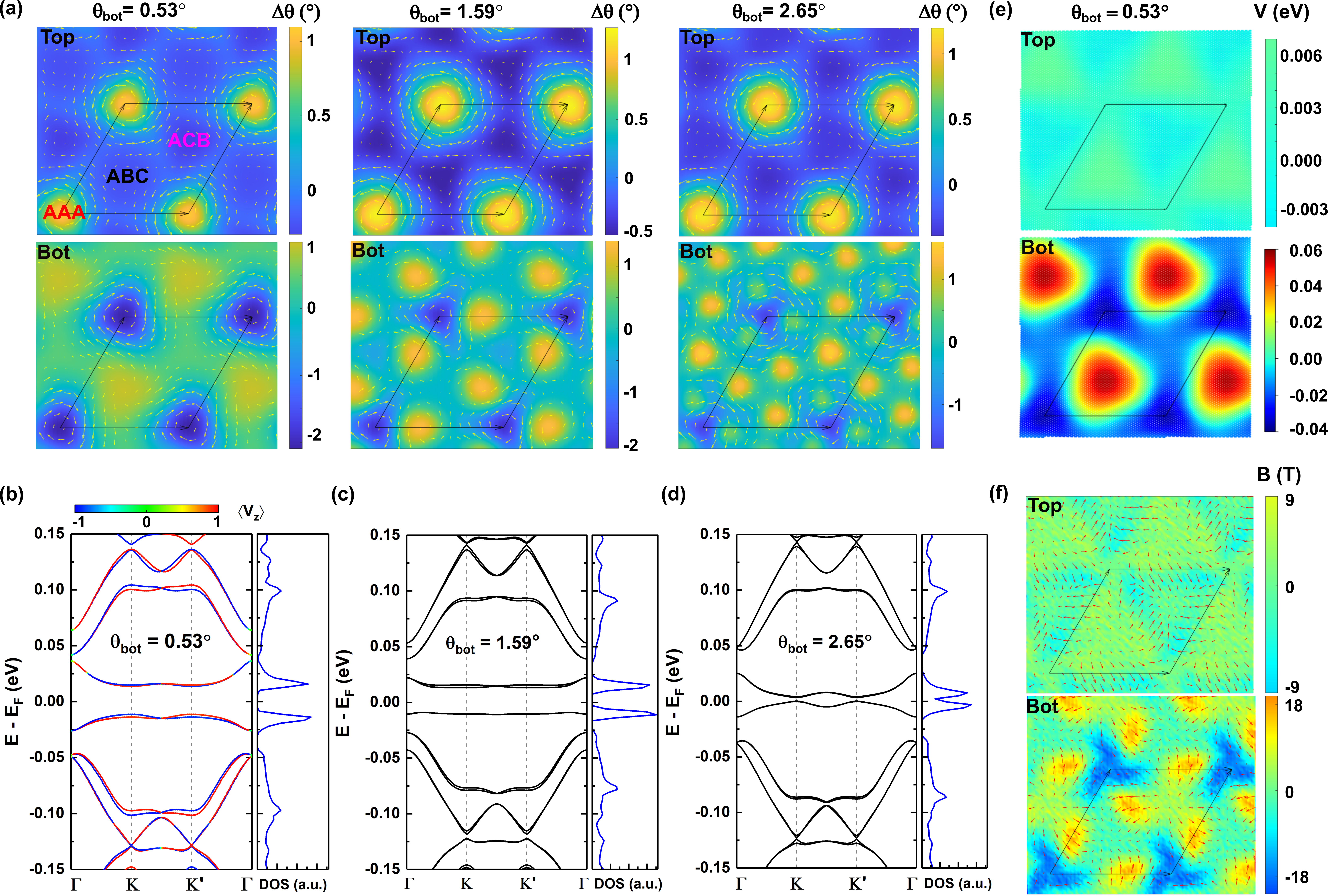}
    \caption{(a) In-plane strain, $\mathbf u(\mathbf r)$, in TBG/hBN with fixed $\theta_{tbg}=1.05^\circ$ and varying $\theta_{bot}$. The in-plane displacements are visualized with white arrows; the color data denotes the local value of the in-plane twist of the atoms with respect to their original position $(\Delta \theta=\nabla \times \mathbf u)$ with positive values corresponding to the counterclockwise rotation. The moir\'{e} supercell is outlined in black. (b) Band structure and density of states of TBG/hBN with $\theta_{bot}=0.53^\circ$. The color bar denotes the band for each valley $\langle \mathbf V_z \rangle$ with $\langle \mathbf V_z \rangle \approx 1$ if a state belongs to valley $K$ and $\langle \mathbf V_z \rangle \approx -1$ if a state belongs to valley $K'$. (c) and (d) Band structure and density of states of  TBG/hBN with $\theta_{bot}=1.59^\circ$ and $\theta_{bot}=2.65^\circ$, respectively. (e) The deformation potential $V_D$ and (f) pseudo-magnetic field $\mathbf B=\nabla \times \mathbf A$ induced by lattice relaxations in the TBG/hBN with $\theta_{bot}=0.53^\circ$. The vector field $\mathbf A(\mathbf r)$ is visualized with red arrows in (f). The twist angle between bilayer graphene is fixed to $\theta_{tbg}=1.05^\circ$ in all cases.}
    \label{fig:potential_band}
    \end{figure*}
    
\section{Substrate induced potentials}\label{sec:subs-ind-pot}

The lattice deformation due to the relaxation gives rise to periodic scalar and gauge potentials within the moir\'e unit cell ~\cite{Jung2017,Sachs2011,Kindermann2012,San-Jose2014,San-Jose2014a,Wallbank2013}. 
In particular, the structural deformation leads to an on-site potential proportional to the local compression/dilatation and a pseudo-vector potential proportional to the shear deformations~\cite{van2015relaxation,Shi2020Nature}. 
All these effects can be accurately considered in the TB calculation in an atomistic level. 
For the on-site potential $V_{D}(\mathbf R_i)\rightarrow V_{D,i}$ of a carbon atom $i$, following Ref.~\cite{slotman2015effect}, we first calculate an effective area, $S_i$, around each $i$-th carbon atom which will be modulated by local deformations, then we obtain the on-site potential at site $i$, with $V_{D,i}$ given by:
\begin{equation}
    V_{D,i}=g_1 \frac{\Delta S_i}{S_{0}},
    \label{eq: deformation_pot}
\end{equation}
where $\Delta S_i = S_i-S_0$, with $S_0=3\sqrt{3}a_0^2/4$ the effective area in equilibrium and $g_1=4$ eV for graphene, which corresponds to the screened deformation potential. This value gives a reasonable description of transport properties~\cite{ochoa2011temperature}, and is close to density functional calculations~\cite{choi2010effects}. In free standing TBG, the deformation potential in Eq.~(\ref{eq: deformation_pot}) originates a periodic scalar on-site potential centered at the unit cell origin~\cite{Shi2020Nature}. As shown in Fig.~\ref{fig:potential_band}(e), the presence of a substrate induces an irregular periodic deformation potential. This irregularity reveals the complexity of the periodic potentials induced by the hBN on TBG which are found to be completely different from those of a single graphene layer on a hBN substrate~\cite{Jung2017}. 
Note that from the Hamiltonian in Eq.~(\ref{eq: TBHamil}), we have two contributions to the on-site energy: the first one is the second term which is due to the energy difference between nitrogen and boron and gives rise to different adhesion energies and is the main source of the mass gap in graphene/hBN heterostructures~\cite{Sachs2011,Kindermann2012,San-Jose2014}, and the second one is the third term that is due to the deformation potential and slightly modifies the band structure (see Supplemental Material ~\cite{SI} Sec. D).
This potential also induces small periodic local mass gaps due to the potential difference between neighbouring sites~\cite{slotman2015effect}.
    
In addition to the periodic scalar potentials, the structural relaxation gives rise to a pseudo-vector potential~$\mathbf{A}$ with components proportional to the shear deformation~\cite{Vozmediano2010,Wijk2015}. In real space, the components of the vector potential are given by:
\begin{align}
    A_x&=\frac{\sqrt{3}}{2} (t_1-t_2), \\ \nonumber
    A_y&=\frac{1}{2} (2t_3-t_1-t_2), 
    \label{eq: A components}
\end{align}
where $t_{i=1,2,3}$ are the first nearest-neighbors inter-atomic interactions in the deformed lattice. If the strain is not uniform ($t_1 \neq t_2 \neq t_3$), the vector potential generates, in general, a pseudomagnetic field given by 
\begin{equation}
     \mathbf{B}=\frac{c}{ev}(\nabla \times \mathbf{A}),
\label{eq: B field}
\end{equation}
where $v$ is the Fermi velocity of graphene and $c=1$ is a numerical factor depending on the detailed model of chemical bonding.
The effect of the pseudomagnetic field is introduced into the Hamiltonian in Eq. (\ref{eq: TBHamil}) by modifying the distance-dependent hopping parameters in Eq.~(\ref{SK_para}). 
The pseudomagnetic field that acts on electrons from the valley $K'$ is exactly opposite to that acts on electrons from the valley $K$. Therefore, time-reversal symmetry is preserved.

\section{Results}\label{sec:results}

\subsection{TBG/hBN structure}

 \subsubsection{Scalar and pseudomagnetic fields}
We first consider TBG on hBN where the geometrical information, including twist angle and stacking, are described in Fig.~\ref{fig:atomic}. The displacements of the atomic sites, the magnitude of the pseudo-magnetic field, the local deformation potentials and the effect of the hBN on the band structure are shown in Fig.~\ref{fig:potential_band} for $\theta_{tbg}=1.05^\circ$ (see Supplemental Material ~\cite{SI} Sec. C) for additional results). Some interesting effects are in order here: for a twist angle $\theta_{bot}=0.53^\circ$, at the corners of the unit cell the atomic displacements in $G_{bot}$ do not have a circular \textquote{vortex} shape as in TBG (see Supplemental Material ~\cite{SI} Sec. B)~\cite{Wijk2015, Jain2016relaxation,Cantele2020Relax}, instead, they have a triangular-like shape. 
As shown in Fig.~\ref{fig:potential_band}(a), the displacements have a counterclockwise behaviour at the corners (AAA stacking configuration in Fig.~\ref{fig:atomic}) and clockwise displacement in the inner regions (ABC and ACB). 
The magnitude of the in-plane twist of the atoms with respect to their original position, $\Delta \theta$, has different values along the moir\'e because the corresponding atomic binding energies are different.
This is due to the different stackings found in each region~\cite{Sachs2011,San-Jose2014}. 
Figure~\ref{fig:potential_band}(e) displays the distribution of the deformation. 
As we can see, in $G_{bot}$ its maximum value is about $60$ meV at the ABC region and a minimum of $-40$ meV is found at the AAA region. 
Notice that in the ACB region the on-site energies are small, as in the AAA centers. This behaviour is reminiscent of that of monolayer graphene on hBN~\cite{Sachs2011,Caciuc2012}. 
In addition, these magnitudes are quite close to those of graphene on a hBN substrate~\cite{slotman2015effect}. 
In $G_{top}$ the on-site potential is one order of magnitude smaller. Figure~\ref{fig:potential_band}(f) shows the vector potential $\mathbf{A}$ (red arrows) and the pseudomagnetic field (in color) induced in both $G_{bot}$ and $G_{top}$. 
The pseudomagnetic field in $G_{bot}$ has a complex shape similar to a \textquote{fidget spinner}, and is completely different to that of monolayer graphene on hBN~\cite{San-Jose2014} or free-standing TBG~\cite{Shi2020Nature}. 
The resulting non-uniform pseudo-magnetic field in $G_{bot}$ has a maximum value of about $18$~T (which is almost twice as in pristine TBG~\cite{Shi2020Nature}). 
In $G_{top}$, the pseudomagnetic field is smaller with a maximum of $9$~T, as shown in  Fig.~\ref{fig:potential_band}(a). 

In recent studies, the effect that a hBN substrate has on TBG is introduced by means of an effective periodic potential acting only on the nearest graphene layer~\cite{Cea2020TBGhbN,Shi2021Conm, Shin2021asy,Mao2021Quasi} with parameters obtained from calculations of a graphene monolayer on hBN. 
As shown in Fig.~\ref{fig:potential_band} our results indicate that this approach is correct, however, a renormalized set of substrate parameters should be considered while describing the periodic potentials. 
We will discuss this in the following sections.

\subsubsection{Evolution of the periodic potentials and mass gap}\label{sec:tbg-hbn-evolution}

\begin{figure}
\centering
\includegraphics[width=0.48\textwidth]{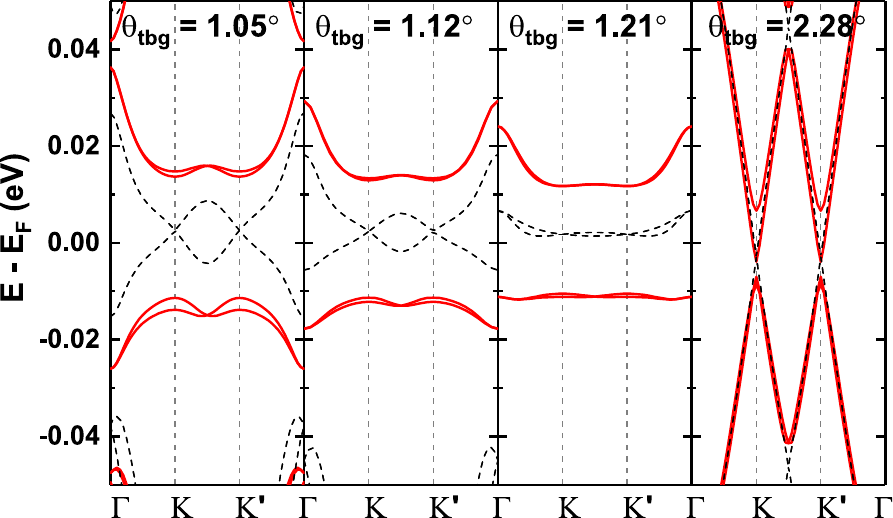}
\caption{Band structure (red solid lines) of TBG/hBN with  $\theta_{bot}=0.53^\circ$ and varying $\theta_{tbg}$. Black dashed lines are the band structures of the corresponding free standing TBG at each $\theta_{tbg}$.}
\label{fig:splitting}
\end{figure}

By increasing $\theta_{bot}$ the effect of the hBN on the TBG band structure is reduced. 
Our results indicate that, for the considered stacking, the effect of the substrate survives for angles larger than $3^\circ$.
%\theta_{bot}\approx3^\circ$.
Nevertheless, it should be pointed out that small variations in the value of $\theta_{bot}$ for angles below that threshold, have a huge impact in the electronic properties of the system.
%Our results indicate that, for the considered stacking, the substrate effects survives for larger angles.
%Nevertheless, if $\theta_{bot}$ is increased to 
As $\theta_{bot}$ increases, shown in Fig.~\ref{fig:potential_band}(a), the period of the moir\'e length between $G_{bot}$ and $hBN_{bot}$ is reduced~\cite{Huang2021Imaging}. 
For example, as detailed in Supplemental Material~\cite{SI} Sec. A, for $\theta_{bot}=1.59^\circ$ we have $q=3$ resulting in $L_{TBG}=3L_{hBN}$. The effect on the band structure is shown in Fig.~\ref{fig:potential_band}(b)-(d), where the narrow bands are separated by a gap resulting from the breaking of inversion symmetry ($\mathcal{C}_2$). 
In the nearly aligned situation, $\theta_{bot}=0.53^\circ$,  the gap between the narrow bands is $\sim 25-30 $ meV. 
As the twist angle of the substrate increases, the second moir\'e length is reduced and the effects of the periodic potentials are suppressed.  For $\theta_{bot}=2.65^\circ$ the gap is nonzero (see Fig.~\ref{fig:potential_band}(c)). The persistence of the gap for angles far from alignment is due to the large value of the mass term resulting from relaxation effects and by the large difference of on-site energies between graphene and hBN sites~\cite{Jung2015,hunt2013experi_butterfly}. 
In addition, as shown in Fig.~\ref{fig:potential_band}(b) and Fig.~\ref{fig:potential_band}(c), the density of states (DOS) for two different angles is quite similar. 
We expect to find similar DOS for smaller values of $\theta_{bot}$. 
The persistence of the band gap for a window of small angles and the isolated narrow bands may explain the presence of quantum anomalous Hall effects (QAH) and other non-trivial band topology effects found in TBG with a nearly aligned hBN substrate~\cite{Setal20,Serlin2019,Sharpe2021orbital}. Our results are consistent with previous studies which argue that the QAH would only appear if one or  both $\theta_{bot}$ and $\theta_{top}$ are near alignment~\cite{Shi2021Conm}.
 
On the other hand, in the trilayer system shown in Fig.~\ref{fig:atomic}, we can tune both the graphene twist angle $\theta_{tbg}$ and the substrate $\theta_{bot}$. As mentioned before, we found that, by increasing $\theta_{bot}$, the effect of the substrate on the TBG electronic properties is reduced. By modifying the graphene twist angle $\theta_{tbg}$ while keeping fixed $\theta_{bot}$, the band structure is also modified. Figure~\ref{fig:splitting} shows the band structure for $\theta_{bot} \approx 0.53^\circ$ and different twist angles, $\theta_{tbg} = 1.05, 1.12, 1.21 \text{ and } 2.28^\circ$.  Here, it is important to note that, in our model, the \textquote{magic} angle is around $\theta_{tbg}\sim 1.21^\circ$. As it can be seen, at this particular angle, the substrate induces a finite gap of about $\approx 30$ meV.  If we decrease the TBG angle, the bands become wider, especially the valence bands. Furthermore, the gap between the valence and conduction bands gap becomes slightly larger. Contrastingly, by increasing the angle, we can see that for $\theta_{tbg}=2.28^\circ$, the AA bilayer graphene band structure is recovered~\cite{Min2008AA,Abdullah2018AA}. 
In this particular case, the presence of hBN has the effect of opening a small gap at the Dirac cones.

\subsubsection{Layer degeneracy breaking}

In free standing TBG, the layer degree of freedom is disentangled from spin and valley, providing eight-fold degeneracy in the low energy states. In the unperturbed system, the Dirac points are at the same energy (dashed black lines in Fig.~\ref{fig:splitting}), and at low energies, Dirac cones only interact with their analogous at opposite layers. The presence of the hBN substrate induces a mass gap. However, the hBN has a larger effect on $G_{bot}$ than on $G_{top}$ and the gapped Dirac cones of $G_{top}$ and $G_{bot}$ are no longer at the same energy. Since the $K$ point corresponding to layer $1$ is mapped to a $K'$ point of layer $2$ in the opposite valley, a \textquote{splitting} in each of the conduction and valence bands is obtained in the full TB model due to the energy difference.  Therefore, this splitting is a breaking of the layer degeneracy because the hBN has a larger effect on $G_{bot}$. In fact, as we will show in the following section, by adding a second hBN layer on the top and for certain twist angles, the degeneracy is recovered. A similar phenomena is obtained in TBG in a uniform electric field~\cite{MoonKoshino2014Interlayer,Yeh2014Interlayer}. 
Next, to identify the bands corresponding to each valley we define a valley operator of the form
\begin{equation}
    \hat{V}_z=\frac{i}{3\sqrt{3}}\sum_{<<i,j>>,s}\eta_{ij}\sigma_z^{ij}c_{i,s}^\dagger c_{j,s},
    \label{eq: valleyop}
\end{equation}
where $i,j$ denotes next-nearest-neighbor sites, $\eta_{ij}=\pm 1$ for clockwise or counterclockwise hopping, respectively and $\sigma_{z}^{ij}$ the Pauli matrix associated with the sublattice degree of freedom. The expectation value of this operator ranges from $-1$ to $+1$, which corresponds to the $K$ and $K'$ valley, respectively. Figure~\ref{fig:potential_band}(b) displays the band structure where the band corresponding to each valley is identified by different colors.

\begin{figure*}
\centering
\includegraphics[width=\textwidth]{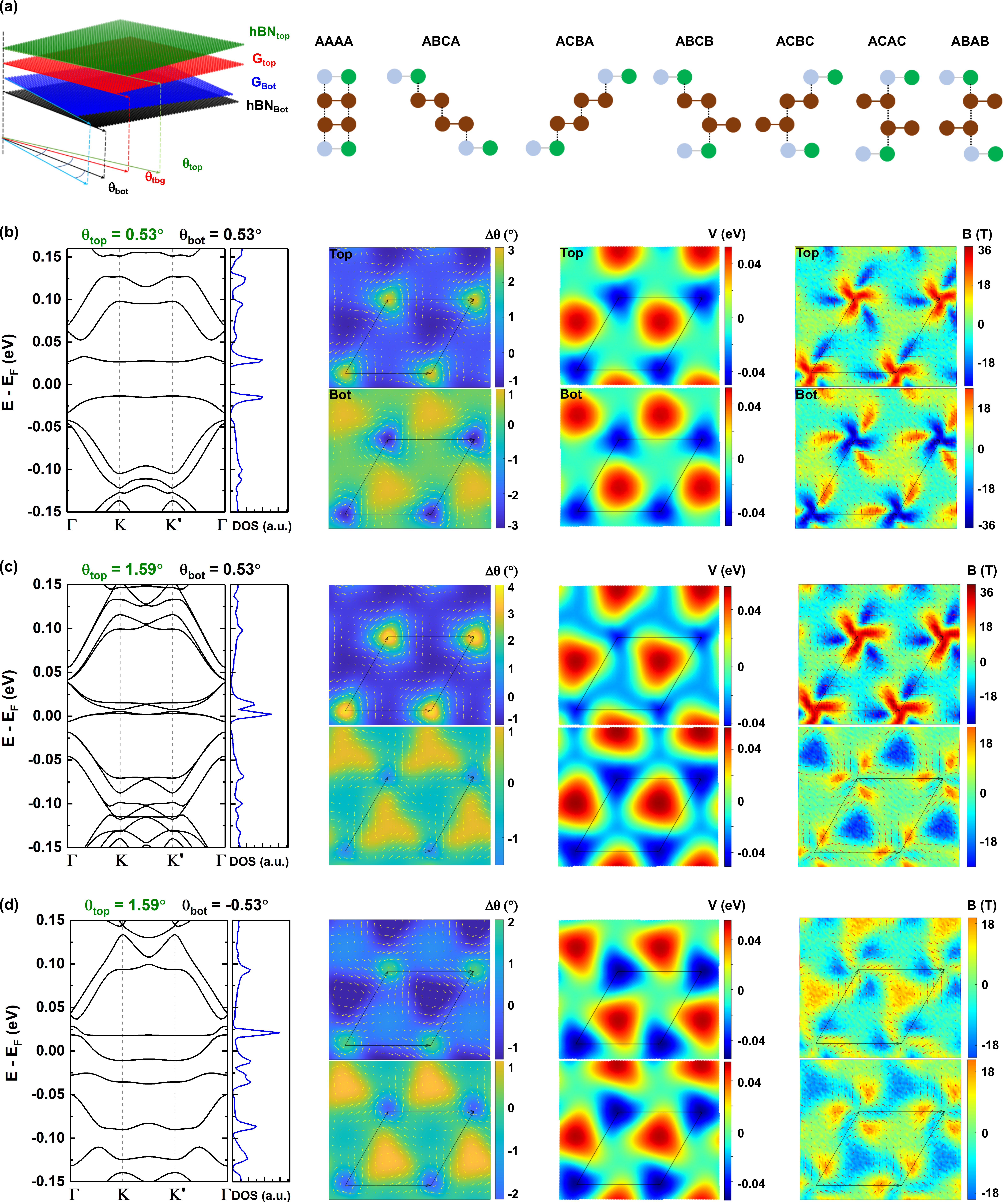}
\caption{(a) Schematic structure of the hBN/TBG/hBN system and the different high-symmetry stackings in the superlattice. (b), (c) and (d) panels from left to right display the band structure, in-plane twist of the atoms with respect to their original position, scalar potential and pseudo-magnetic field. The twist angle of TBG is fixed to $1.05^\circ$.}
\label{fig:sandwich}
\end{figure*}

\subsection{hBN/TBG/hBN structure}

\subsubsection{Scalar and pseudomagnetic fields}

We now consider the case of TBG encapsulated between two layers of hBN. 
Depending on the orientation and twist angle of each hBN layer with respect to TBG, several configurations can be obtained. 
Similar to the trilayer case, our starting point is an AAAA stacking configuration at the unit cell origin. Additional configurations can be obtained by modifying the stacking, see for example Refs. \cite{Jiseon2021TBGhBN,lin2021misalignment}. 
In Fig.~\ref{fig:sandwich}(a) we show a structure with arbitrary $\theta_{bot}$, $\theta_{top}$ and $\theta_{tbg}$, where the graphene layers are depicted in red and blue and the hBN substrate in green and black. 
Furthermore, we also show the different stacking configurations that can be found at the high symmetry points along the moir\'e superstructure.

Due to the three different degrees of freedom to build this system, the number of different heterostructures that can be generated are almost limitless.
Here we are going to focus on three different cases where we fix $\theta_{tbg}$ to $1.05^\circ$ and modify both $\theta_{bot}$ and $\theta_{top}$. 
In Figs.~\ref{fig:sandwich}(b)-(d) we show the corresponding band structure, strain fields, deformation potential and pseudomagnetic fields for the different studied configurations. 
Particularly, for these structures, the high-symmetry stackings are (AAAA, ABCA, ACBA), (AAAA, ABCB, ACBC) and (AAAA, ACAC, ABAB), respectively. 

In Fig.~\ref{fig:sandwich}(b) we show the case when $\theta_{bot}=\theta_{top}$.
It is important to note that, since all the twist angles are measured with respect to $G_{bot}$, the twist angle with respect to $G_{top}$ is $\theta'_{top}=-0.53^\circ$.
Therefore, the lattice structure is such that $hBN_{top}$ and $hBN_{bot}$ are aligned with respect to each other. 
In this case, a large gap of about $\sim 50$~meV is found in the band structure. 
The corresponding pseudomagnetic field has a three-lobed structure with a maximum magnitude of about $36$~T, which is larger than in the trilayer case. 
The magnitude and direction of this field is opposite in each graphene layer and, as shown in the right panel of Figs.~\ref{fig:sandwich}(b)-(d) is highly sensitive to the hBN twist angle. 
In the second column of Fig.~\ref{fig:sandwich}, we can see how the in-plane twist of the atoms with respect to their original position, $\Delta \theta$, has a complex shape which also strongly depends on the hBN angle. 
Interestingly, in $G_{top}$ the in-plane strain field is oriented in the opposite direction with respect to the field in $G_{bot}$. 
The deformation potential is proportional to the change in the effective area around each atom with respect to its equilibrium position, Eq.~(\ref{eq: deformation_pot}). 
%As shown in the third column in Fig.~\ref{fig:sandwich} and considering the parallelogram unit cell, the potential at the corners is smaller and negative with a maximum in one of the interior zones. 
As shown in the third column of Fig.~\ref{fig:sandwich}, the potential at the corners of the unit cell is negative and has a smaller absolute value than those in the interior regions, where there is a maximum.
This behaviour is reminiscent of monolayer graphene on hBN where is know that the moir\'e pattern in the rigid system smoothly changes between $AA$ type, $AB$ type (carbon on boron) and $BA$ type (carbon on nitrogen). Each of these configurations have a different adhesion energy which is minimum at the AB stacking, while BA and AA are roughly similar~\cite{Sachs2011,San-Jose2014,San-Jose2014a}. The different energies create in-plane forces which tends to maximize the area of the AB regions and minimizing the other regions. This behaviour is completely capture by our results in the third column in Fig.~\ref{fig:sandwich} where in the red regions the local area around each atomic site was maximised and by Eq.~(\ref{eq: deformation_pot}) the on-site potential is positive. Interestingly, this color difference allow us to distinguish the different stack configurations between each TBG layer and its neighboring hBN layer by simply looking for the maximum or minimum in the corresponding deformation potential. 

\subsubsection{Layer degeneracy}
Interestingly, in the tetralayer structure hBN/tBG/hBN, for twist angles where $\theta_{bot}=\theta_{top}$,  Fig.~\ref{fig:sandwich}(b), the eightfold degeneracy is recovered. 
As we will elucidate in the following section, this occurs because the periodic potentials acting in each graphene layer are opposite or with the same magnitude. 
On the contrary, if $\theta_{bot}\neq\theta_{top}$, as shown in Fig.~\ref{fig:sandwich}(c), the effect of the substrate is different in each graphene layer and the layer degeneracy is broken resulting in a band splitting. 
Notice that, the pseudomagnetic fields have different maximum values on each layer. 
The same happens for the in-plane strain. 
In principle, we can assume that the layer degeneracy is broken in structures with different $\theta_{bot}$ and $\theta_{top}$, however, this is not always the case. Fig.~\ref{fig:sandwich}(d) shows the TBG band structure with different twist angles $\theta_{bot}$ and $\theta_{top}$ where the layer degeneracy is recovered. By simple geometry we notice that the twist angle of $hBN_{top}$ with respect to $G_{top}$ is $\theta_{top}=0.54^\circ$ which means that each hBN layer is twisted in opposite directions with the same angle with respect to its nearest graphene layer and, therefore, the corresponding pseudomagnetic and strain fields, Fig.~\ref{fig:sandwich}(d), have opposite directions and equal magnitudes. 
Interestingly, the density of states close to charge neutrality has a single peak. 
This indicates that even if $\theta_{bot}$ and $\theta_{top}$ are small (or both hBN layers are nearly aligned with its corresponding graphene layer), the band structure can be modified completely.
This can have important effects in experimental measurements.
The strong dependence of the electronic properties of TBG for small values of $\theta_{bot/top}$ may explain why in some experiments the valley anomalous Hall conductivity in suspended or encapsulated structures of TBG with hBN is not always present~\cite{Setal20,Serlin2019,Tschirhart2021}. 
In addition, our results indicate that adding a single hBN layer to TBG breaks the layer degeneracy giving rise to a splitting in the band structure. This effect will produce a double peak in the local density of states similar to the effect produced by a magnetic field~\cite{Shi2020Nature}. 
%Interestingly, this layer degeneracy can be recovered with a second hBN layer placed on the top.  

\begin{figure*}
\begin{centering}
 \includegraphics[scale=0.245]{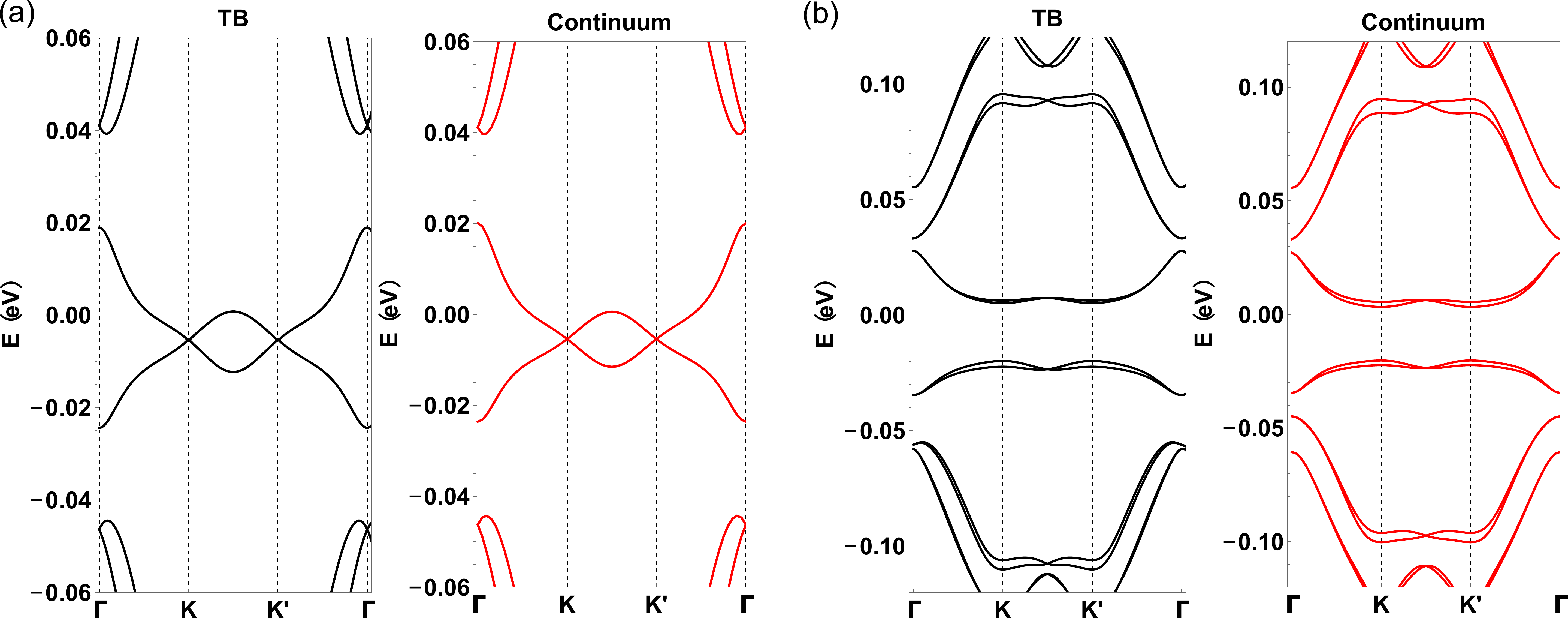} 
\end{centering}
\caption{Band structure of twisted bilayer graphene with $\theta =1.05^\circ $ calculated with TB (black lines) and continuum (red lines) model. (a) free standing TBG and (b) TBG/hBN.  
\label{fig:Fitcont}}
\end{figure*}

\section{Effective continuum model}\label{sec:cont-mod}

As we have seen previously, TB is a powerful tool to study the electronic properties of realistic TBG and related structures in combination with semi-classical relaxation methods. Nevertheless, another common approach in the investigation of TBG is the derivation of a continuum model~\cite{Bistritzer2011} that describes the bands obtained from the TB calculations. This model could help us elucidate some of the obtained effects. Therefore, this section is going to be dedicated to the study of a continuum model describing a system of this work. To do so, we start by numerically fitting the bands of free standing TBG with a continuum model for each valley. After obtaining the correct interlayer parameters and Fermi velocity we will include the effect of a single hBN layer by means of an effective periodic potential.

\subsection{Pristine TBG}

We first consider the case of pristine TBG, where we need to fit four parameters: the velocity term, $v_{f}$, which is known that depends on the twist angle~\cite{LopesDosSantos2007,Trambly2012,Yin2015lc,Oshiyama2015}, the interlayer coupling parameters, $\{u, u'\}$, which strongly depends on the relaxation, and an additional periodic scalar potential $V_s$. The low energy Hamiltonian of TBG is given by, 

\begin{equation}
H=\left(\begin{array}{cc}
H(\boldsymbol{ q_{1,\zeta}})+V_\text{s}(\boldsymbol{r}) & U(\boldsymbol{r})^{\dagger}\\
U(\boldsymbol{r}) & H(\boldsymbol{ q_{2,\zeta}})+V_\text{s}(-\boldsymbol{r})
\end{array}\right),
\label{eq:HamiltonianTBG}
\end{equation}
where $\boldsymbol{ q}_{l,\zeta}=R(\pm\theta/2)(\boldsymbol{ q}-\boldsymbol{ K}_{l,\zeta})$ with $\boldsymbol{ K}_{l,\xi}$ the graphene Dirac cones and $H(\boldsymbol{ q})=-(\hbar v_{f}/a)\boldsymbol{ q}\cdot(\zeta\sigma_{1},\sigma_{2})$ is the Hamiltonian for monolayer graphene and $\zeta=\pm1$ is a valley index. In the above equation, $U(\boldsymbol{r})$ is the interlayer coupling between graphene layers which is given by the Fourier expansion, 

\begin{align}
U & =\left(\begin{array}{cc}
u & u^{\prime}\\
u^{\prime} & u
\end{array}\right)+\left(\begin{array}{cc}
u & u^{\prime}\omega^{-\zeta}\\
u^{\prime}\omega^{\zeta} & u
\end{array}\right)e^{i\zeta\boldsymbol{ G_1}\cdot\boldsymbol{ r}}+ \\ \nonumber 
&+ \left(\begin{array}{cc}
u & u^{\prime}\omega^{\zeta}\\
u^{\prime}\omega^{-\zeta} & u
\end{array}\right)e^{i\zeta(\boldsymbol{ G_1}+\boldsymbol{ G_2})\cdot\boldsymbol{ r}},
\label{eq: InterlayerU}
\end{align}
where $\omega=e^{2\pi i/3}$, with $u$ and $u^{\prime}$ the amplitudes which take into account relaxation effects as described in Ref.~\citep{Koshino2018a,TKV19,Nam2017}. $\boldsymbol{ G}_{1}$ and $\boldsymbol{ G}_{2}$ are the reciprocal lattice vectors. 
We also introduce an even and periodic scalar potential,
\begin{equation}
V_s\left(\boldsymbol{r}\right)=-V_0 \sum_{j=1}^{6} e^{i\boldsymbol{G_j}\cdot\boldsymbol{r}},
\label{eq: vopot}
\end{equation}
with $V_0$ a constant term and the sum running over the six nearest neighbors reciprocal lattice vectors (or fist star), this is $\{\boldsymbol{{G}_1,{G}_2,{G}_3,-{G}_1,-{G}_2, -{G}_3 \}}$, with $\boldsymbol{{G}_3= -({G}_1+{G}_2)}$. 
We found that, although the magnitude of $V_0$ is small, it has to be included in order to correctly fit the continuum model to the TB results.
We believe that this potential is related to the different atomic rearrangements at the AA, AB and BA sites due to the relaxation~\cite{Oshiyama2015,Yin2015gpot,Rademaker2018}.

In Fig.~\ref{fig:Fitcont} we show the fittings that we obtained using the parameters in Table~\ref{tb:FittingValues} which are in agreement with the typically accepted values for TBG~\citep{Nam2017,Guinea2019CModels}.
As expected, our results clearly indicate that the continuum model parameters strongly depend on the twist angle. 
The scalar potential appears to increase with the twist angle, however, for large angles, it becomes negligible. 
This potential, which is even and periodic, slightly distorts the narrow bands in a similar way as a Hartree potential with a negative filling would~\cite{Cea2019Pinning}. 

\begin{table}
\begin{centering}
\begin{tabular}{cccccccccc}
\hline 
\hline 
Twist angle  && $\hbar v_{f}/a$ && $u$ (meV) && $u'$ (meV) && $V_{0}$ (meV)\\%\tabularnewline
\hline 
$1.05^{\circ}$ && $-2.25$ && $63.3$ && $120$ && $-1.6$\tabularnewline
%\hline 
$1.12^{\circ}$ && $-2.31$ && $68.6$ && $120$ && $-1.8$\tabularnewline
%\hline 
$1.20^{\circ}$ && $-2.34$ && $71.7$ && $119.3$ && $-2.2$\tabularnewline
%\hline 
$1.30^{\circ}$ && $-2.20$ && $67.0$ && $110.7$ && $-2.5$\tabularnewline
%\hline 
$1.35^{\circ}$ && $-2.09$ && $61.2$ && $104.4$ && $-2.7$\tabularnewline
\hline 
\hline
\end{tabular}
\par\end{centering}
\caption{Values obtained by fitting the continuum model to the TB parameters.}
 \label{tb:FittingValues}
\end{table}

\subsection{TBG on hBN}

As stated in Section \ref{sec:tbg-hbn-evolution}, adding a hBN substrate to support a graphene monolayer breaks the inversion symmetry ($\mathcal{C}_2$) of the crystal which results in gapped Dirac cones~\cite{Hunt2013,SL13,Amet2013,Getal14,Cetal14,Yankowitz2014,Wetal15,Jung2015,Lee_science2016,Wetal16,Yankowitz_nat2018,Zibrov_natphys18,Kim2018}.
By placing TBG on hBN, two coexisting moir\'e patters arise: the first one induced by the relative orientation between the two graphene layers, and the second appearing due to the lattice constant mismatch between hBN and the bottom graphene layer~\cite{Xue2011,Yankowitz2012b,Woods2014,Moon2014,San-Jose2014}.
To qualitatively fit a continuum model to our TB results, we  assume that the coexisting moir\'e patterns are identical. 
Previous studies~\cite{Cea2020TBGhbN,Shi2021Conm,Mao2021Quasi,lin2021misalignment,Shin2021asy} have considered the effect of an hBN substrate on TBG by means of continuum models that use the parameters obtained by TB models of graphene on hBN~\cite{Xue2011,Yankowitz2012b,Woods2014,Moon2014,San-Jose2014}. 
Only in Ref. \cite{Shin2021asy} a vertical relaxation within the continuum model is taken into account.
To our knowledge, there are no previous works where a full TB model for TBG on hBN that takes into account lattice relaxations is used to fit such models.
%To our knowledge, there are no previous works where a full TB model for TBG on hBN is used with the lattice relaxations taken into account. 
It is important to note that, although the fitting of the continuum model is qualitative, the full TB that we use is widely accepted and is in agreement with first-principles calculations as well as it has been used to support different experimental results~\cite{Shi2020Nature,Anelkovi2020,Huder2018}.

\subsubsection{Effective potential and fitting parameters}

The effect of the substrate can be introduced as an effective potential acting on the closest graphene layer which is periodic in the moir\'e unit cell~\cite{Wallbank2013,San-Jose2014}:
\begin{eqnarray}
V_\text{SL}\left(\boldsymbol{r}\right)= \omega_0 \sigma_{0}+
\Delta\sigma_{3}+
\sum_{j}
V_\text{SL}(\boldsymbol{G_j})e^{i\boldsymbol{G_j}\cdot\boldsymbol{r}},
 \label{eq: VSlhbN}
\end{eqnarray}
with amplitudes $V_\text{SL}(\boldsymbol{G_j})$ given by
\begin{equation}
V_\text{SL}(\boldsymbol{G_j})= V_{s}(\boldsymbol{G_j})+V_{\Delta}(\boldsymbol{G_j})+V_{g}(\boldsymbol{G_j}),
\label{eq: hbN Perturbation}  
\end{equation}
where
\begin{align}
V_{s}(\boldsymbol{G_j})&= \left[V_{s}^{e}+i(-1)^{j}V_{s}^{o}\right]\sigma_{0},\\ \nonumber V_{\Delta}(\boldsymbol{G_j})&=\left[V_{\Delta}^{o}+i(-1)^{j}V_{\Delta}^{e}\right]\sigma_{3}, \\ \nonumber
V_{g}(\boldsymbol{G_j})&=\left[V_{g}^{e}+i(-1)^{j}V_{g}^{o}\right]M_j,\\ \nonumber
\end{align}
with $M_j=(-i\sigma_{2}G_{j}^{x}+i\sigma_{1}G_{j}^{y})/|G_{j}|$. The $2 \times 2$ Pauli matrices act on the sublattice index in a single graphene layer. 
$\Delta$ is a parameter that represents a spatially uniform mass term~\cite{Hunt2013}. 
Surprisingly, having identified that some of the potential profiles in the TB solutions indicate the existence of additional harmonics (see Supplemental Material ~\cite{SI} Sec. G), we found that the modulation of $V_\text{SL}$ at smaller wavelengths has some effects on the narrow bands and, therefore, Eq.~(\ref{eq: VSlhbN}) has to be expanded up to two harmonics. 
As described in Ref.~\cite{Wallbank2013,San-Jose2014}, the minimal model for a hBN substrate depends on eight parameters: uniform on-site term $\omega_0$, constant mass gap $\Delta$, position-dependent scalar terms $V_{s}^e$ and $V_{s}^o$ which are even and odd under spatial inversion, respectively. 
Two position dependent mass terms, $V_{\Delta}^o(e)$ and two gauge terms $V_{g}^o(e)$, odd and even respectively. 
In the case of the second harmonic, six additional periodic potential parameters are required.  
Therefore, the Hamiltonian of TBG/hBN depends on several parameters, four for TBG (Eq.~(\ref{eq:HamiltonianTBG}) and Table~\ref{tb:FittingValues}) and eight for the substrate, Eq.~(\ref{eq: VSlhbN}). 
The exact values and their dependence with the twist angles are known for single layer graphene on hBN~\cite{Jung2017} but they are still unknown for the combined system TBG/hBN or hBN/TBG/hBN. 
Obtaining exact values for a large set of parameters is a difficult task since several combinations can give similar results and a complete fitting is out of the scope of this work. 
In addition, it is important to note that, the induced periodic potentials and mass gap due to the presence of a hBN substrate dominate over $V_0(r)$ of the free standing system (Eq.~(\ref{eq: vopot})), which, in this particular case, is in fact negligible.

Some qualitative estimations of the substrate parameters can be given by analyzing the band structures and the periodic potentials in Fig.~\ref{fig:potential_band}. 
For the considered stacking in Fig.~\ref{fig:atomic}, an estimated set of parameters is given by 
$(w_{0},\Delta,V_{1,s}^{e},V_{1,s}^{o},V_{1,\Delta}^{e},V_{1,\Delta}^{o},V_{1,g}^{e},V_{1,g}^{o})= (3.0, 31.62, -0.75, 0.68,-0.02, 3.4, -5.14, 18.6)$ meV, for the first harmonic amplitudes and the only non-zero amplitudes for the second harmonic are the gauge fields $(V_{2,g}^{e},V_{2,g}^{o})= (-5.14, -9.3)$.  
Fig.~\ref{fig:Fitcont}(b) shows the band structure with the full TB model (black line) and the fitting with a continuum model (red line) using our estimated set of parameters which agree qualitatively with the TB calculations (see Supplemental Material ~\cite{SI} Sec. H). 
In particular, in order to achieve this kind of agreement, we found that: i) the mass gap is large and ii) the gauge terms require up to two harmonics. 
The first feature is in agreement with Ref.~\citep{Jiseon2021TBGhBN}, where the band structure with a similar stacking has a large mass gap. 
As shown in Fig.~\ref{fig:potential_band}, as $\theta_{bot}$ increases, the corresponding moir\'e length is reduced. 
This strongly suppresses the effect of the periodic potentials induced by the hBN and only the constant mass terms survives for large angles. 
As a final note, the strong gauge field in ii) results from the effects of the relaxation of the trilayer system (TBG/hBN). 

\subsubsection{Layer degeneracy breaking: mass gap}

\begin{figure}
    \centering
    \includegraphics[scale=0.28]{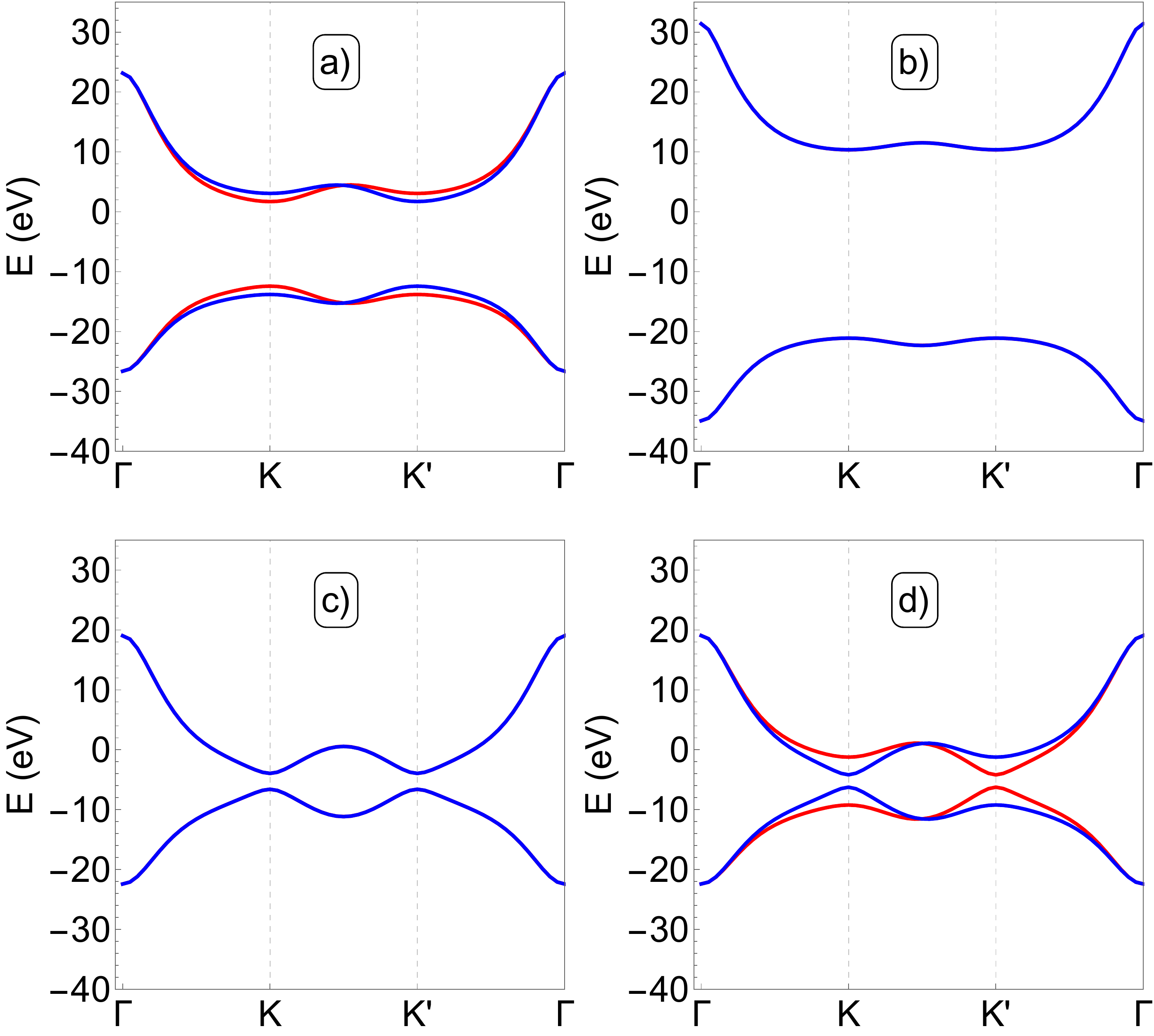}
    \caption{Band structure of pristine TBG with different mass terms. a) $\Delta_b=30$ meV and $\Delta_t=0$, (b) $\Delta_b=\Delta_t=30$ meV, (c) $\Delta_b=-\Delta_t=30$ meV and (d) $\Delta_b=30$ meV and $\Delta_t=-40$ meV. In each panel, blue and red lines are the bands corresponding to each valley. The periodic potentials are set to zero.}
    \label{fig:massgap}
    \end{figure}

By fitting the continuum model to the TB band structures we have found that the dominant terms are the gauge fields which give rise to a pseudomagnetic field and the mass term which is responsible for the large gap. 
In the following, for simplicity, we only consider the effects of the mass term. 
A similar analysis can be performed with the gauge potential. 
Fig.~\ref{fig:massgap} shows the band structure of TBG in both valleys with only the constant mass term $\Delta$ of Eq.~(\ref{eq: VSlhbN}). 
We define $\Delta_b$ and $\Delta_t$ as the mass terms acting in $G_{bot}$ and $G_{top}$ respectively. Figs.~\ref{fig:massgap}(a) and (d) clearly show the band splitting due to the different mass acting on each graphene layer. 
On the contrary, Figs.~\ref{fig:massgap}(b) and (c) are still degenerated due to the fact that the mass terms for the bottom and top layers are the same. 
This explains why in some of our TB results the presence of two hBN layers does not split the narrow bands. 
Our results indicate that in an experimental setup, if a hBN layer is nearly aligned with its closest graphene layer, the layer degeneracy is broken, resulting in a double peak in the local density of states, similar to the effect produced by a magnetic field~\cite{Shi2020Nature}. 
We would like to recall that, in the encapsulated case, depending on the orientation of the hBN layers the degeneracy can be recovered. 

%%%%%%%%%%%%%%%%%%%%%%%%%%%%%%%%%
\section{Conclusions}
In this work we have studied the electronic properties of TBG supported on or encapsulated in hBN via a combination of an accurate real-space TB model and semi-classical molecular dynamics.
This procedure allows us to have a good description of realistic structures that are measured in experiments. Using this approach, we have found that hBN affects the electronic properties of TBG even when the angle between TBG and hBN is far from 0$^\circ$.
When studying TBG supported on a hBN substrate, the band structure shows a gap that separates the flatbands which we ascribe to a mass gap induced by the hBN.
Moreover, hBN induces pseudomagnetic fields which, in combination with the mass terms break the layer degeneracy and splittings within the bands appear.
Interestingly, when adding an extra hBN layer, that is, when the TBG is clamped between two hBN layers the gap between the flat bands still appears, although the degeneracy of the bands can be recovered for certain twist angles. Finally, we have also developed a continuum model that correctly describes the calculated TB electronic properties. This kind of models are helpful to understand the underlying physics behind the main features of the TBG/hBN heterostructures. We would like to stress that, in order to describe realistic structures as the ones found in experimental devices, our approach of mixing semi-classical molecular dynamics with an atomistic tight-binding model is nowadays a state-of-the-art calculation since first-principles calculations are far from feasible due to the number of atoms in the structures. 
Therefore, the explanation of novel phenomena appearing in TBG either supported on a hBN or embedded between two of those layers such as superconductivity, correlated insulators or the recently found ferroelectric phase might only be possible using an accurate TB calculation or continuum models that describe the electronic properties of such systems accurately.

\section*{Acknowledgements}
This work was supported by the National Science Foundation of China (Grants No.11774269 and No.12047543).
IMDEA Nanociencia acknowledges support from the ``Severo Ochoa" Programme for Centres of Excellence in R\&D (Grant No. SEV-2016-0686).
P.A.P and F.G. acknowledge funding from the European Commission, within the Graphene Flagship, Core 3, grant number 881603 and from grants NMAT2D (Comunidad de Madrid, Spain), SprQuMat. S.Y. acknowledges funding from the National Key R\&D Program of China (Grant No. 2018YFA0305800). Numerical calculations presented in this paper have been performed in the Supercomputing Center of Wuhan University.  We are thankful to Tommaso Cea for many productive conversations.

% and SEV-2016-0686 (Ministerio de Ciencia e Innovación, Spain).  

\bibliographystyle{apsrev4-1}
\bibliography{citations_maintext.bib}

\end{document}